\documentclass[aps,preprint,showpacs]{revtex4}
\usepackage{graphics}                           
\usepackage{epsfig}                             
\usepackage{dcolumn}
\begin{document}
\preprint{New format}

\title{Shell structure at $N=28$ 
near the dripline:  spectroscopy of $^{42}$Si, $^{43}$P and $^{44}$S} 

\author{J. Fridmann$^1$, I. Wiedenh\"over$^1$, A. Gade$^2$, L.T. Baby$^1$, 
D. Bazin$^2$, B.A. Brown$^2$, C.M. Campbell$^2$, J.M. Cook$^2$, P.D. Cottle$^1$, 
E. Diffenderfer$^1$, D.-C. Dinca$^2$, T. Glasmacher$^2$, P.G. Hansen$^2$, 
K.W. Kemper$^1$, J.L. Lecouey$^2$, W.F. Mueller$^2$, 
E. Rodriguez-Vieitez$^3$, J.R. Terry$^2$, J.A. Tostevin$^4$, K. Yoneda$^2$,
and H. Zwahlen$^2$}

\affiliation{$^1$Department of Physics, Florida State University, 
Tallahassee, FL  32306-4350\\
$^2$National Superconducting Cyclotron Laboratory and 
Department of Physics and Astronomy, Michigan State University, 
East Lansing, MI  48824-1321, USA \\
	$^{3}$Nuclear Science Division, Lawrence Berkeley 
National Laboratory, Berkeley, CA  94720, USA \\
	$^4$Department of Physics, School of Electronics 
and Physical Sciences, University of Surrey, Guildford, Surrey GU2 7XH, 
United Kingdom\\}


\date{\today}

\begin{abstract}

Measurements of the $N=28$ isotones $^{42}$Si, $^{43}$P and $^{44}$S using 
one- and two-proton knockout reactions from the radioactive beam nuclei 
$^{44}$S and $^{46}$Ar are reported.  The knockout reaction cross sections
for populating $^{42}$Si and $^{43}$P and a 184 keV $\gamma$-ray observed in
$^{43}$P establish that the $d_{3/2}$ and $s_{1/2}$ proton orbits are
nearly degenerate in these nuclei and that there is a substantial $Z=14$ subshell
closure separating these two orbits from the $d_{5/2}$ orbit.  The increase in the 
inclusive two-proton knockout cross section from $^{42}$Si to $^{44}$S demonstrates 
the importance of the availability of valence protons for determining the cross
section.  New calculations of the 
two-proton knockout reactions that include diffractive effects are presented.
In addition, it is proposed that a search for the $d_{5/2}$ 
proton strength in $^{43}$P via a higher
statistics one-proton knockout experiment could help determine the size of the
$Z=14$ closure.

\end{abstract}

\maketitle

\section{Introduction}

The quest for information about shell structure in $N=28$ nuclei near the neutron 
dripline is of central importance to the field of nuclear structure physics for 
two reasons.  First, the nuclei in the vicinity of $^{42}$Si provide the first 
arena in which ideas about how changes in the spin-orbit force affect 
shell structure near the 
neutron dripline can be tested.  
The $N=28$ shell closure is the lightest neutron shell 
closure caused by the spin-orbit force, which is responsible for all shell 
closures in heavier nuclei.  
At present, the $N=28$ shell closure 
is the only major neutron shell closure driven by the spin-orbit force that is 
experimentally accessible.
It has been predicted that the spin-orbit 
force affecting the neutron orbits weakens close to the neutron dripline 
both because of the weak binding of neutrons in this vicinity and the 
role of the continuum \cite{Na01,Wa04}.  Indeed, it has been predicted 
\cite{We94,We96,Te97,La98,La99,Pe00,Ro02} that the $N=28$ shell closure should 
be less well developed, or even collapse altogether, in the nuclei near 
$^{42}$Si.  Three recent experimental results, one a measurement of 
the lifetime of the $\beta$-decay of $^{42}$Si \cite{Gr04}, the second the 
determination that $^{43}$Si is bound \cite{No02} and the third a mass
measurement of $^{42}$Si \cite{Sa05} have been used to argue 
that the $N=28$ 
shell closure has narrowed or collapsed in 
$^{42}$Si, resulting in a well-deformed shape 
for this nucleus.  On the other hand, it has been argued in Refs. 
\cite{Co98,Ca04} that a possible proton subshell closure at $Z=14$ would have 
a strong effect on the structure of $^{42}$Si, preventing it from being 
well-deformed.   

The second reason that experiments on nuclei near $^{42}$Si are important is 
that they are providing a rigorous testing regime for experimental techniques 
that will be used heavily at the next generation of radioactive beam facilities, 
including the Rare Isotope Accelerator.  Among these techniques are the 
intermediate-energy knockout reactions in which cross sections provide 
spectroscopic information similar to that obtained for many years from direct 
transfer reactions used at low-energy stable beam facilities 
\cite{Ha03,Ba03}.

In the present article, we provide a comprehensive report of a set of 
measurements of the $N=28$ isotones $^{42}$Si, $^{43}$P and $^{44}$S using 
one- and two-proton knockout reactions from the radioactive beam nuclei 
$^{44}$S and $^{46}$Ar.  A brief account of this work was given in Ref. 
\cite{Fr05}.  In addition to more detail regarding the results in \cite{Fr05}, 
this article also reports:  new calculations of the two-proton knockout 
reactions that include diffractive contributions that were not available prior to 
the publication of \cite{Fr05}; new shell model calculations that allow us 
to refine the conclusions we draw from the two-proton knockout results 
regarding the nature of the $Z=14$ shell closure; and calculations of the 
expected distribution of $d_{5/2}$ proton hole strength in the one-proton 
knockout reaction spectrum of $^{43}$P.  

Section II of this article will include details of the experimental 
techniques and results.  Section III will discuss the experimental results 
on $^{42}$Si and $^{44}$S and the theoretical calculations on two-proton knockout reactions.  
In Section IV, we discuss the experimental and theoretical results regarding 
$^{43}$P.  Section V will 
provide a brief summary.

\section{Experimental Details}

The present experiments were performed at the National Superconducting 
Cyclotron Laboratory at Michigan State University using the Coupled Cyclotrons 
Facility (CCF).  Beams of the radioactive nuclei $^{44}$S and $^{46}$Ar were 
produced via fragmentation of a primary beam of 140 MeV/nucleon $^{48}$Ca 
provided by the CCF.  The primary beam was fragmented on a 705 mg/cm$^2$ 
thick beryllium target, and the fragmentation products were separated in the 
A1900 fragment separator \cite{A1900}.  The separator selected secondary 
beams of 98.6 MeV/nucleon $^{44}$S and 98.1 MeV/nucleon $^{46}$Ar.  The 
$^{44}$S secondary beam had a momentum spread of $\pm 1.0\%$ and a 
purity of $75\%$; the rate of $^{44}$S particles impinging on the secondary 
target averaged 400 particles/s.  For the $^{46}$Ar secondary beam, the 
momentum spread was $\pm 0.5\%$ and the purity was $95\%$; the beam 
rate on the secondary target was 240,000 $^{46}$Ar particles/s.  Both the 
$^{44}$S and $^{46}$Ar secondary beam particles were identified event-by-event
by their time-of-flight before impinging on the secondary target.   

The knockout reactions were induced on a secondary beryllium target of 
thickness 375 mg/cm$^2$.  The residual projectile-like nuclei were detected in 
the S800 spectrograph \cite{S800}.  Gamma-rays emitted at the secondary 
target location were detected using the SeGA array of segmented high 
purity germanium detectors \cite{SeGA} in coincidence with the 
residues in the S800 spectrograph.

The measurement of the residual nucleus $^{42}$Si was performed via the 
two-proton knockout reaction on the $^{44}$S secondary beam,
$^9$Be($^{44}$S,$^{42}$Si)X.  The integrated 
secondary beam was $1.14 \times 10^8$ $^{44}$S particles.  
The $^{43}$P measurement 
was performed with the one-proton knockout reaction on the $^{44}$S secondary 
beam, $^9$Be($^{44}$S,$^{43}$P)X; the integrated secondary 
beam for this measurement was $1.1 \times 10^7$ $^{44}$S 
particles.  Finally, $^{44}$S was measured via the two-proton knockout 
reaction on the $^{46}$Ar secondary beam, $^9$Be($^{46}$Ar,$^{44}$S)X 
with an integrated secondary beam of 
$7.3 \times 10^9$ $^{46}$Ar particles.  

The spectra from the S800 used to identify residual nuclei are shown in Fig. 1.  
The vertical axes correspond to the energy loss in the ion chamber at 
the focal plane of the S800, while the horizontal axis plots the 
path-corrected time of flight 
between the object point of the spectrograph and the focal plane.  
The inclusive cross sections for production of the 
$^{42}$Si, $^{43}$P and $^{44}$S residual nuclei were 0.12(2) mb, 7.6(11) mb 
and 0.23(2) mb, respectively.  The momentum distributions for these residual
nuclei were within the acceptance of the S800 spectrograph for the respective
reactions and S800 settings.

The $\gamma$-ray spectra in coincidence with the $^{42}$Si, $^{43}$P and 
$^{44}$S residual nuclei are shown in Fig. 2.  
These spectra are Doppler-reconstructed
so that they appear as in the rest frames of the residual nuclei.  The total 
photopeak efficiency of the SeGA array for this experiment was $5.7\%$ at 180 keV,
$2.2\%$ at 1 MeV, and $1.3\%$ at 2 MeV.  

There are no discernable $\gamma$-ray peaks in the $^{42}$Si spectrum.

The $^{43}$P spectrum includes a single large peak at 184(3) keV.  No other 
peaks are clearly discernable, although there are background 
counts up to 1 MeV.  

For $^{44}$S, the $2_1^+ \rightarrow 0_{gs}^+$  $\gamma$-ray previously reported in
Refs. \cite{Gl97,So02,Gr05} is seen clearly in our spectrum and we assign
an energy of 1.329(10) MeV.  This $\gamma$-ray was first observed in the Coulomb 
excitation study of Glasmacher {\it et al.} \cite{Gl97}, 
where an energy of 1.297(18) MeV was
given.  More recently, Sohler {\it et al.}, in a report of their
study of $^{44}$S via the fragmentation of $^{48}$Ca \cite{So02}, gave an energy of
1.350(10) MeV for this transition.  Finally, the report of the observation
of an isomer in $^{44}$S in Ref. \cite{Gr05} is accompanied by the report 
of an energy of 1.329 MeV for the $2_1^+ \rightarrow 0_{gs}^+$ transition,
with no experimental uncertainty given.  It is likely there are other $\gamma$-rays
in the spectrum below 1.2 MeV, but the resolving power of the present experiment does
not allow us to distinguish them clearly.

\section{$^{42}$Si and $^{44}$S}

\subsection{Shell structure}

The present measurement of $^{42}$Si provided two experimental 
conclusions: first, that the inclusive 
cross section for producing this nucleus in the 
two-proton knockout reaction is 0.12(2) mb, and second, that there are no 
discernable $\gamma$-rays in the spectrum.  

The inclusive cross section is small, smaller than any previously observed 
for the two-proton knockout reaction \cite{Ba03}.  Bazin {\it et al.} \cite{Ba03} 
pointed out that the cross section for the two-proton knockout reaction 
depends on the number of valence protons, so the small cross section observed 
here suggests that a shell closure occurs at $Z=14$ – where the $d_{5/2}$ 
proton orbit fills.  Indeed, in the $N=28$ isotone $^{48}$Ca, the 
($d$,$^3$He) reaction \cite{Do76,Ba85} revealed a large gap between the 
$d_{5/2}$ proton orbit and the $d_{3/2}$ and $s_{1/2}$ proton orbits 
(which are nearly degenerate in $^{48}$Ca).  
The small two-proton knockout cross section 
provides strong evidence that the $Z=14$ shell gap is substantial in $^{42}$Si 
as well.

We have performed shell model calculations and reaction calculations to 
examine how neutron and proton shell structure affect the 
spectroscopy of $^{42}$Si and $^{44}$S.  
The shell model calculations use the interaction 
of Ref. \cite{Nu01} with the neutrons occupying a model space including the 
$0f_{7/2}$ and $1p_{3/2}$ orbits and the protons occupying the full $sd$ space.  
The two-nucleon amplitudes that resulted were used to calculate two-proton 
knockout cross sections with a model that extends that described in Ref. 
\cite{Fr05} by including diffractive effects.  

\newcommand {\la} {\langle}
\newcommand {\ra} {\rangle}
\newcommand {\beq} {\begin{eqnarray}}
\newcommand {\eeqn} [1] {\label{#1} \end{eqnarray}}
\newcommand {\eol} {\nonumber \\}
\renewcommand {\vec} [1] {\mbox{\boldmath $#1$}}

\subsection{Two-Proton Removal Calculations}

The two-proton removal reaction from an intermediate energy
neutron-rich projectile nucleus has recently been shown to proceed
as a sudden, direct reaction process \cite{Ba03,Tos04}. While the
associated two-proton structures and the reaction mechanism were
treated only approximately in Ref.\ \cite{Ba03}, a far more
complete calculational scheme has since been detailed in Ref.\
\cite{Tos04}. This dealt with that part of the two-nucleon removal
cross section arising from the stripping (inelastic breakup)
mechanism and the theoretical approach combined fully the
two-nucleon shell-model configurations and their associated
spectroscopic amplitudes with eikonal direct reaction theory. This
analysis provided further evidence for the direct nature of the
reaction mechanism in such systems. The theoretical calculations
of the two-proton-removal cross sections presented here follow the
formalism developed in Ref.\ \cite{Tos04}. In addition,
we include a full calculation of the cross section contributions
from the mechanism in which one proton is absorbed (stripped)
while a second is elastically dissociated (diffracted) by the
target. An estimate of the (smaller) cross section due to the
removal of both nucleons by the elastic dissociation (diffraction)
mechanism is also included, as is outlined below.

The two knocked-out nucleons are assumed to be removed from a set
of active, partially occupied single-particle orbitals $\phi_{j}$
with spherical quantum numbers $n(\ell j)m$. The shell model wave
function of the removed nucleons in the projectile ground state,
relative to any given residue (or core) state $f$, is the
sum over the contributing two-particle configurations,
\begin{eqnarray}
\Psi_{J_iM_i}^{(f)}(1,2)= \sum_{I \mu \alpha} C_\alpha^{J_i J_f
I}(I\mu J_f M_f|J_i M_i) [\overline{\phi_{j_1}\otimes
\phi_{j_2}}]_{I\mu} ~. \label{bigone}
\end{eqnarray}
Here $\alpha$ denotes these available configurations ($j_1,j_2$)
and $[\overline{\phi_{j_1}\otimes \phi_{j_2}}]$ is their
normalized, antisymmetrized wave function \cite{Tos04}. The
$C_\alpha^{J_i J_f I}$ are the signed two-nucleon spectroscopic
amplitudes which are calculated here using the shell model code
{\sc oxbash} \cite{Oxbash}.

The model used for the two-nucleon stripping cross section was
discussed fully in \cite{Tos04}, to which the reader is referred.
The partial cross section to a residue final state $f$ is the
integral over all projectile center-of-mass (cm) impact parameters
$b$ and average over the two-removed-nucleon wave functions
\begin{equation}
\sigma_{str}= \frac{1}{(2J_i+1)} \sum_{M_i} \int d\vec{b}\, |{\cal
S}_c|^2\, \langle \Psi_{J_i M_i}^{(f)}| (1-|{\cal S}_1|^2)
(1-|{\cal S}_2 |^2)| \Psi_{J_iM_i}^{(f)}\rangle ~, \label{sum}
\end{equation}
where the ${\cal S}_i$ are the eikonal $S$-matrices \cite{Tos01}
for the elastic scattering of the two nucleons (1,2) and the
$A$-body core with the target. These are functions of their impact
parameters and are assumed to be spin-independent. This cross section
expression accounts for those events in which the residue emerges 
from the reaction having missed or interacted only elastically 
with the target, as described by $|{\cal S}_c|^2$, and two nucleons are
removed through inelastic collisions with the target.  This inelasticity
and the associated removal of flux from the nucleon-target elastic
channels is described by the product of the nucleon-target absorption
probabilities $(1-|{\cal S}_i|^2)$.

Additional contributions to the knockout cross section, when one
nucleon, say 1, is removed in an elastic interaction with the
target while the second nucleon is absorbed, enter the eikonal
model expression for the absorption cross section via the term
\begin{equation}
\sigma_{1}=\frac{1}{2J_i+1}\sum_{M_i} \int d\vec{b}\, |{\cal
S}_c|^2 \,\langle \Psi_{J_i M_i}^{(f)}| |{\cal S}_1 |^2
\,(1-|{\cal S}_2|^2)| \Psi_{J_i M_i}^{(f)}\rangle ~, \label{sum1}
\end{equation}
and similarly for nucleon 2. These diffraction-plus-absorption
terms require further attention since the cross section in Eq.\
(\ref{sum1}) includes processes in which nucleon 1 remains bound
to the residue. These correspond to a single nucleon absorption
from the projectile populating bound states of an ($A$+1)-body
residue. Such effects could be ignored in an analogous discussion
of the nuclear breakup of Borromean nuclei, such as $^{11}$Li
\cite{BandE}, where there are no two-body bound (valence) states
of the core and the (non-absorbed) neutron. These single-nucleon
stripping contributions are removed by projecting off bound
nucleon-residue final states, by replacing
\begin{eqnarray}
|{\cal S}_1|^2 \rightarrow {\cal S}_1^* \LARGE[1-\sum_{j''m''}
|\phi_{j''}^{m''} ) ( \phi_{j''}^{m''} |\, \LARGE] {\cal S}_1~,
\label{complete}
\end{eqnarray}
in Eq.\ (\ref{sum1}). Here the notation implies a summation over
the bound eigenstates $n(\ell'' j'')m''$ of  nucleon 1 and the
core and we have used the $(..|$ and $|..)$ bra-kets to denote
integration over the coordinates of this single nucleon. In the
calculations presented we include all the active single particle
orbitals in this sum. We find that the $\sigma_{1}+\sigma_2$
contribution to the cross section is similar to $\sigma_{str}$.

Finally, we include an estimate of the (smaller) cross section due
to the removal of both (tightly-bound) nucleons by elastic
dissociation. Our estimate makes use of the reduction in the cross
section when a single nucleon is elastically dissociated compared
to it being stripped, $\sigma_{i}/\sigma_{str}$. We thus estimate
the two-nucleon elastic breakup cross section to be $\sigma_{diff}
\approx [\sigma_{i} /\sigma_{str}]^2 \sigma_{str}$. Since, for the
cases discussed here, and more generally, $\sigma_{i} /
\sigma_{str} \approx 0.35 - 0.4$, then $\sigma_{diff}$ makes a
contribution of order $6-8\%$ to the two-proton removal partial
cross sections. The theoretical two-proton removal cross sections
presented are the sum of these contributions, i.e. $\sigma_{th} =
\sigma_{str}+ \sigma_{1} + \sigma_2 + \sigma_{diff}$.

The $S$-matrices in Eq. (\ref{sum}) were calculated from the core
and target one-body matter densities using the optical limit of
Glauber's multiple scattering theory \cite{Gla59,Tos01}. A
Gaussian nucleon-nucleon (NN) effective interaction was assumed
\cite{Tos99} with a range of 0.5 fm. This calculates residue- and
nucleon-target $S$-matrices and corresponding reaction cross
sections in line with measurements in the 50-100 MeV/nucleon
energy range, e.g. \cite{Kox87}. The strength of the interaction
was determined, in the usual way \cite{Ray79}, by the free pp and
np cross sections and the real-to-imaginary ratios of the forward
NN scattering amplitudes, $\alpha_{pp}$ and $\alpha_{np}$. The
latter affect the calculation of the diffractive contributions but
are of no consequence for the stripping terms, that require only
$|{\cal S}_i|^2$. For the $^9$Be($^{46}$Ar,$^{44}$S)$X$ and
$^9$Be($^{44}$S,$^{42} $Si)$X$ reaction calculations the density
of $^9$Be was assumed to be of Gaussian form with a root mean
squared (rms) matter radius of 2.36 fm \cite{Oza01}. The densities
of the core nuclei, $^{44}$S and $^{42}$Si, were taken from
spherical, Skyrme (SkX interaction) Hartree-Fock calculations
\cite{skx}. These have rms matter radii of 3.45 fm and 3.44 fm,
respectively.

Our calculations suppose that the removed protons occupy the
$0d_{5/2}$, $0d_{3/2}$ and $1s_{1/2}$ sd-shell orbitals. The
nucleon single-particle wave functions were calculated in a
Woods-Saxon potential well with the conventional radius and
diffuseness parameters $r_0=1.25$ fm and $a=0.70$ fm. The
strengths of the binding potentials were adjusted to support bound
eigenstates with half the physical ground-state to ground-state
two-proton separation energies. These two-proton separation
energies were taken as $S_{2p}=33.42$ MeV and 40.46 MeV for
$^{46}$Ar and $^{44}$S. Also,
due to these large separation energies, we did not included the
small corrections to the nucleon separation energies for
transitions to bound, excited final states.

Full details of the formalism for these new diffractive 
terms, and their application to several (test-case) $sd$-shell 
nuclei, with better-understood structures, will be presented
elsewhere \cite{ToXX}. Using the full $sd$-shell model spectroscopy,
the calculated inclusive cross sections consistently overpredict
the measured cross sections by a factor of two, requiring a
corresponding suppression,
$R_s(2N) \approx 0.5$, of the two-nucleon shell model 
strengths. The analogous suppression effects in nucleus-
induced \cite{Ha03,Ga04b} and electron-induced \cite{Kr01} single-nucleon 
knockout reactions are now well documented, and are of 
order $R_s(1N) \approx 0.6$ for a wide range of nuclei with ($N,Z$) asymmetries 
and/or separation energies similar to those of the present 
study.

\subsection{Discussion}

The calculated inclusive two-proton knockout cross section for $^{42}$Si, 
using wavefunctions generated with the shell-model effective interaction of 
Ref. \cite{Nu01}, is 0.32 mb, about a factor of three larger than
the experimental value of 0.12(2) mb.  However, taking into account the
expected $R_s(2N)$ value of order 0.5, discussed above, there is a closer
agreement with the measured value.  It is also worth noting 
that $92\%$ of the cross
section calculated for $^{42}$Si using these parameters is located in the
ground state; therefore, the contribution of the excited states to the inclusive
cross section measurement is likely to be small. 

As mentioned above, the reason
that the cross section is small is because of the $Z=14$ shell gap;
that is, the energy gap between the $d_{3/2}$-$s_{1/2}$ pair and the
$d_{5/2}$ orbit is large.  We can examine how 
the theoretical cross section depends on the size of the $Z=14$ gap
by performing several calculations in which the gap is reduced.  The 
Ref. \cite{Nu01} parametrization sets the $d_{3/2}-d_{5/2}$ gap as 5.9 MeV.
There is an experimental reason to believe that this gap may be too
large - an analysis of the centroids of proton hole strength observed
in $^{47}$K via the $^{48}$Ca($d$,$^3$He) reaction and reported in
\cite{Ba85} gives a $d_{3/2}$-$d_{5/2}$ gap of 4.8 MeV.  
  
More calculations were performed in which this gap was reduced by
1 MeV and 3 MeV.  In addition, the size of the $N=28$ neutron gap was also 
reduced by 1 MeV (from its value of 3.6 MeV for
$^{42}$Si) to examine how this affects the
cross section.  In all, four calculations were performed - the first
with both the proton and neutron gaps at the values from Ref. \cite{Nu01}; 
the second
with the neutron gap reduced by 1 MeV and the proton gap left at
the value from Ref. \cite{Nu01}; the third with both the proton and neutron gaps
reduced by 1 MeV from the Ref. \cite{Nu01} values; and the fourth with the neutron
gap reduced by 1 MeV from the Ref. \cite{Nu01} value and the proton gap reduced by
3 MeV from the \cite{Nu01} value.

The cross section results from the four calculations,
including the $R_s(2N)=0.5$ suppression, are shown in the
top panel of Fig. 3.  
The reduction of the neutron gap does not affect
the two-proton knockout cross section.  Furthermore, the reduction of
the proton gap by 1 MeV does not significantly affect the cross section, 
either.  This is not surprising since even with this reduction the gap
is 4.9 MeV.  However, a 3 MeV reduction in the proton gap does result in
a significant increase in the cross section.
  
The bottom two panels of Fig. 3 show two other sets of spectroscopic results
for $^{42}$Si
from the four shell model calculations - the energy of the lowest $2_1^+$
state, $E(2_1^+)$, and the reduced electromagnetic matrix element connecting
the $2_1^+$ state to the ground state, $B(E2;2_1^+ \rightarrow 0_{gs}^+)$.
Of these two observables, the reduced matrix element is the more reliable
indicator of quadrupole collectivity.  Reducing the neutron gap by 1 MeV
has a strong effect on both these observables.  Adding the 1 MeV reduction 
of the proton gap has little additional effect on $E(2_1^+)$, but causes
a significant additional increase in $B(E2;2_1^+ \rightarrow 0_{gs}^+)$.  
The large (3 MeV) reduction in the proton gap from its original value of
5.9 MeV causes a near-doubling in the $B(E2;2_1^+ \rightarrow 0_{gs}^+)$
matrix element from its value with a 1 MeV proton gap reduction.  At this
point, proton excitations are playing a large role in driving deformation.

A calculation of the inclusive two-proton knockout cross section for $^{44}$S
using shell model wavefunctions determined with 
the parameters of Ref. \cite{Nu01} yields a result of 0.66 mb, which is
(as in the case of $^{42}$Si) much larger than the experimental value
of 0.23(2) mb.  Once again, the $R_s(2N)$ systematics lead to an expected
theoretical value of around 0.33 mb, in closer agreement with the measured value.
The calculation qualitatively reproduces the increase in
cross section from $^{42}$Si to $^{44}$S with the addition of valence protons.

It is reasonable to conclude from comparing the cross section data to these 
calculations that there is a large gap at $Z=14$, although these data 
cannot provide a quantitative measure of the size of this gap.  With respect
to the urgent question of whether the $N=28$ gap has narrowed from its size
in $^{48}$Ca or even disappeared altogether, the present data cannot provide 
any insights.  Instead, a measurement of $B(E2;2_1^+ \rightarrow 0_{gs}^+)$ 
in $^{42}$Si would provide much more information on the size of the neutron gap and,
therefore, the strength of the orbital spitting between $l=1$ and $l=3$ and spin-orbit 
interaction on neutron orbits near the neutron dripline.

\section{$^{43}$P}

The one-proton knockout reaction preferentially populates states
that have the structure of a proton hole in the beam nucleus \cite{Ha03}.  
Therefore, the
states in $^{43}$P populated with the largest cross sections in the 
one-proton knockout reaction on $^{44}$S are expected to be those that 
are single protons in the $d_{3/2}$ or $s_{1/2}$ orbits, or a single 
$d_{5/2}$ proton hole.  In $^{47}$K, (a proton hole coupled to the
doubly-magic nucleus $^{48}$Ca), the centroids of the strength from the
$d_{3/2}$ and $s_{1/2}$ proton orbits are seen to be separated
by only 300 keV \cite{Do76,Ba85}.  In the present measurement of $^{43}$P, 
the 184 keV $\gamma$-ray suggests that the strength of one of these two
single-proton orbits is concentrated in the ground state with the strength
of the other orbit concentrated in an excited state at 184 keV.  The
$d_{5/2}$ strength is expected at higher excitation energies in $^{43}$P,
as discussed below.

The cross sections for the ground state and 184 keV state are large and 
therefore support this picture.  The inclusive cross section - which
includes both the ground state and the 184 keV state - is 7.6(11) mb.
An examination of the residue-$\gamma$-ray coincidences shows that the
184 keV state accounts for $75 \pm 15 \%$ of the cross section.  The 
combination of this
observation regarding the relative cross sections of the two states and 
calculations
based on the prescription given in \cite{Tos01,Ga04} provide a strong argument that
the $s_{1/2}$ proton strength is concentrated in the ground state, while the
$d_{3/2}$ strength is concentrated in the 184 keV state.  A shell model
calculation similar to that described in section III using the Ref. 
\cite{Nu01} parameters
(including the 5.8 MeV $d_{3/2}-d_{5/2}$ proton orbit splitting), yields a
cross section of 3.7 mb for a ground state that consumes $98\%$ of the $s_{1/2}$
strength, while a state that consumes $99\%$ of the $d_{3/2}$ strength is
located at 0.20 MeV and has a cross section of 7.9 mb.  The experimental and theoretical
inclusive cross sections are thus in the ratio $R_s=0.66(9)$, consistent with
systematics of suppression factors from other well-bound-nucleon knockout
studies \cite{Ha03,Ga04}.  The excited state fraction from theory, $68\%$, is
also in good agreement with the measured value of $75(15)\%$.  
If the $d_{3/2}-d_{5/2}$ proton orbit splitting is reduced
by 1 MeV (to reflect the experimental result in $^{47}$K as discussed in 
section III), then the ground state holds $95\%$ of the $s_{1/2}$ strength
and has a cross section of 3.5 mb, with $98\%$ of the $d_{3/2}$ strength 
residing in
a state at 0.24 MeV that has a 7.7 mb cross section.  These two calculations
are consistent with each other in that they predict that the $d_{3/2}$
state has a cross section a little more than twice the $s_{1/2}$ cross section,
which is approximately what is seen in the data.

As demonstrated in Section III, the spin-orbit splitting of the 
$d_{3/2}$ and $d_{5/2}$ proton orbits is an important parameter for 
interpreting the results of the present
two-proton knockout 
study of $^{42}$Si, other measurements of this nucleus and data on nearby nuclei.
While no evidence for $d_{5/2}$ strength was observed in the present data set,
the one-proton knockout reaction on $^{44}$S 
provides a means for determining the $d_{3/2}-d_{5/2}$ proton spin-orbit splitting.  
A calculation of the distribution of $d_{5/2}$ proton hole strength in $^{43}$P 
using the Ref. \cite{Nu01} parameters, including the 5.8 MeV $d_{3/2}-d_{5/2}$  
splitting and the resulting cross sections for population of these states 
in the one proton knockout reaction (with the cross section calculation 
prescription as described in Refs. \cite{Tos01,Ga04}) is shown in the top
panel of Fig. 4.  It provides a prediction of a 
concentration of $d_{5/2}$ proton hole strength (7.2 mb of cross section) at 
2.24 MeV (although it is quite likely this strength is fragmented over several 
states - explaining its apparent absence from the spectrum of Fig. 2), 
with a somewhat smaller concentration (2.2 mb) at 1.54 MeV.  This 
yields a centroid of 2.1 MeV.  If instead the $d_{3/2}-d_{5/2}$ splitting is 
set to 4.8 MeV, as shown in the bottom panel of Fig. 4, 
the centroid of the $d_{5/2}$ strength is 1.5 MeV 
(with a 6.4 mb concentration at 1.67 MeV and a 7.4 mb concentration at 1.32 
MeV).  We conclude that the location of the $d_{5/2}$ proton hole strength in the 
one-proton knockout reaction provides a sensitive scale for determining the 
$d_{3/2}-d_{5/2}$ proton spin-orbit splitting, and that this provides a 
motivation for another $^{44}$S one-proton knockout experiment with the sensitivity
(increased by means of greater statistics) required to detect the fragments 
of the $d_{5/2}$ strength.

\section{Summary}

The present data on the $N=28$ isotones $^{42}$Si, $^{43}$P and $^{44}$S provide
important insights regarding the most important factors influencing nuclear 
structure in this vicinity.  First of all, the knockout reaction cross sections
for populating $^{42}$Si and $^{43}$P and the 184 keV $\gamma$-ray observed in
$^{43}$P firmly establish that the $d_{3/2}$ and $s_{1/2}$ proton orbits are
nearly degenerate in these nuclei and that there is a substantial $Z=14$ subshell
closure separating these two orbits from the $d_{5/2}$.  The increase in the 
inclusive two-proton knockout cross section from $^{42}$Si to $^{44}$S demonstrates 
the importance of the availability of valence protons for determining the cross
section.  In addition, a search for the $d_{5/2}$ 
proton strength in $^{43}$P via a higher
statistics one-proton knockout experiment could quantify the size of the
$Z=14$ closure.

\begin{acknowledgments}

This work was supported by the National Science Foundation through grants
PHY-0456463, PHY-0110253, PHY-0244453 account 61-2275 (7032), the Department
of Energy through grant DE-FG02-02ER-41220, the State of Florida,
and by the United Kingdom Engineering and Physical Sciences Research 
Council (EPSRC) Grant No. EP/D003628

\end{acknowledgments}

\begin{figure}
\epsfig{file=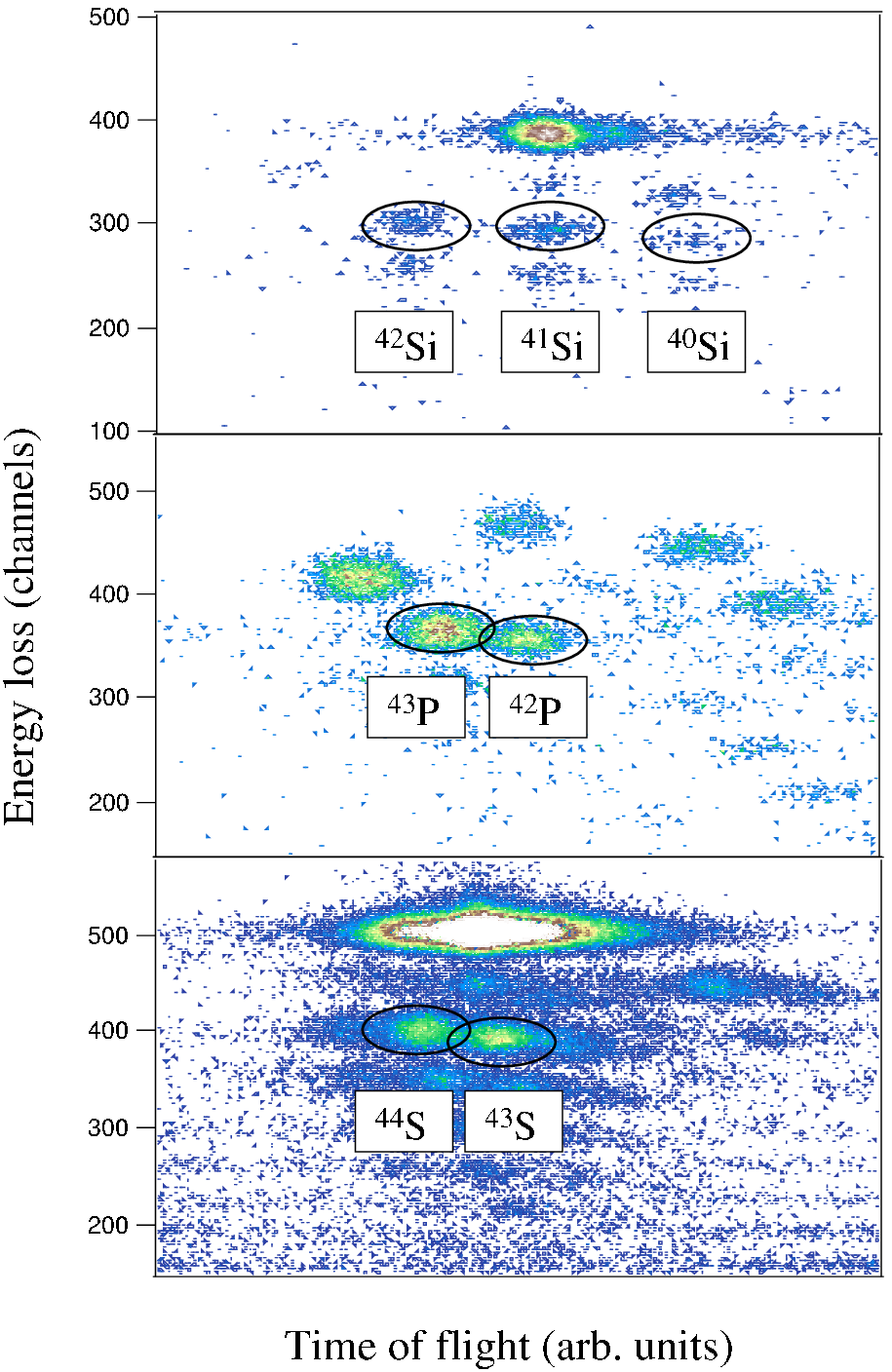,height=5in,angle=0}
\caption{Particle spectra used to identify $^{42}$Si from the two-proton
knockout reaction from $^{44}$S (top), $^{43}$P from the one-proton knockout
reaction from $^{44}$S (middle), and $^{44}$S from the two-proton knockout
reaction from $^{46}$Ar (bottom).  The energy loss in the ion chamber
at the focal plane of the S800 spectrograph is plotted on the vertical axis, 
and the horizontal axis plots the path-corrected time of flight 
between the object point of the spectrograph and the focal plane, with 
shorter flight times to the right.}
\end{figure}

\begin{figure}
\epsfig{file=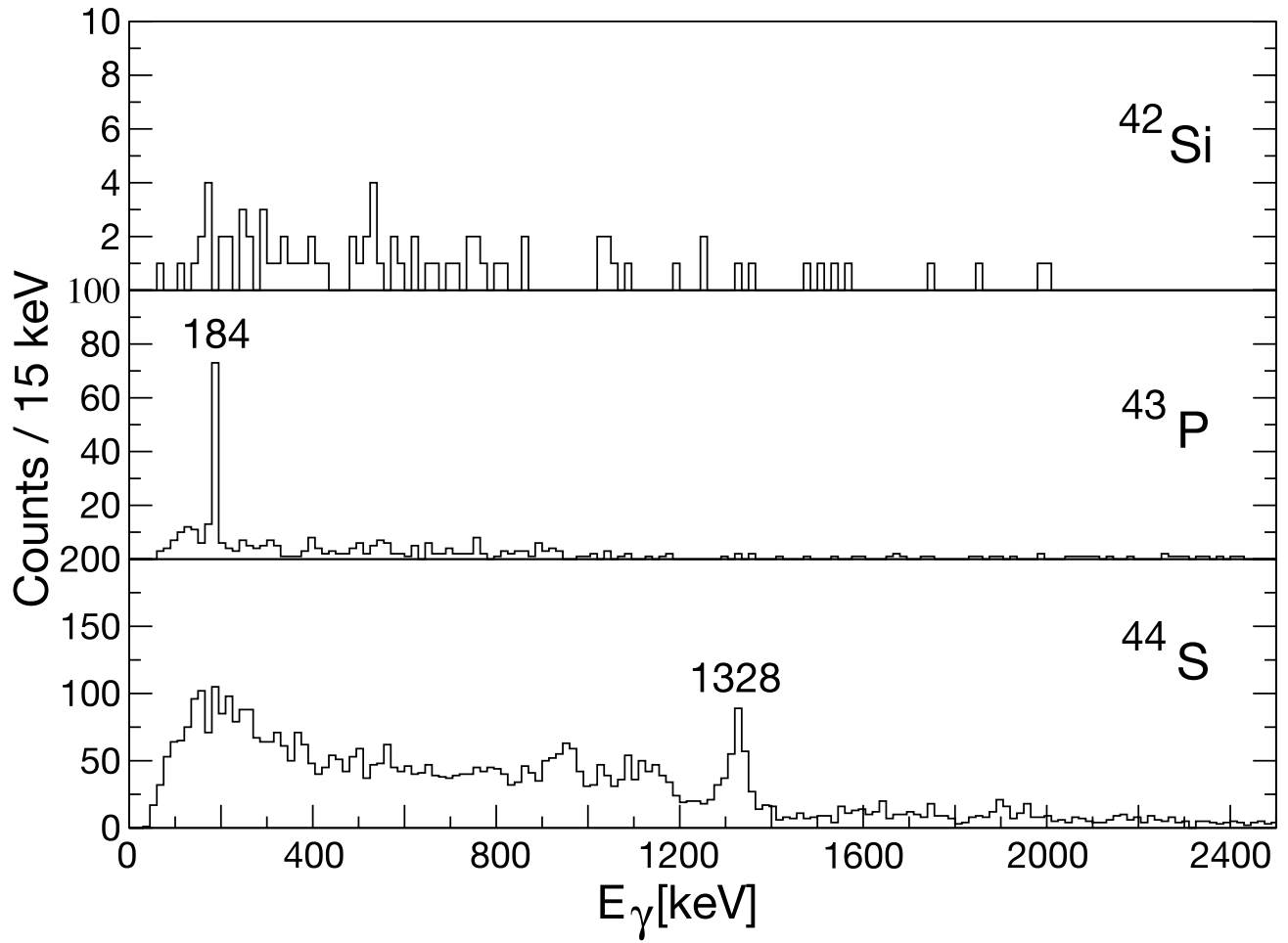,height=5in,angle=0}
\caption{Spectra of $\gamma$-rays detected in coincidence with $^{42}$Si 
(top panel), $^{43}$P (middle) and $^{44}$S (bottom) in the 
$^{44}$S($^9$Be,$^{42}$Si)X, $^{44}$S($^9$Be,$^{43}$P)X, and 
$^{46}$Ar($^9$Be,$^{44}$S)X reactions, respectively.}
\end{figure}

\begin{figure}
\epsfig{file=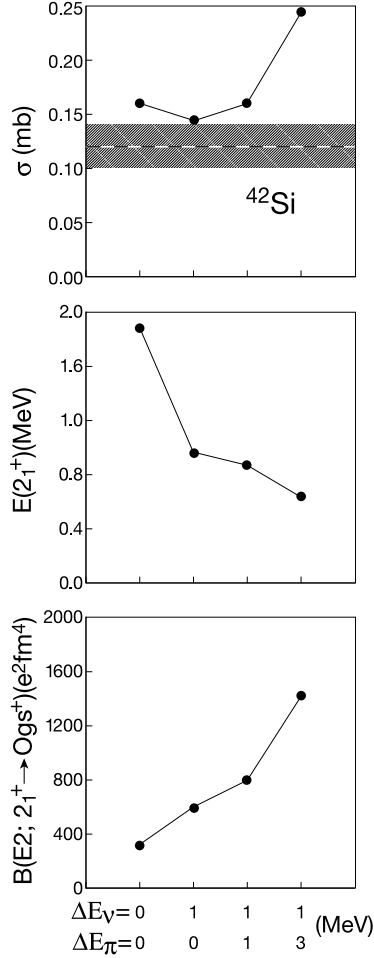,height=5in,angle=0}
\caption{Spectroscopic observables in $^{42}$Si calculated with the
shell model and four sets of parameters for the $Z=14$ and $N=28$ (sub)shell
closures, as described in the text.  The four sets of parameters are denoted by
$\Delta E_{\nu}=0, \Delta E_{\pi}=0$ (shell gaps used in Ref. \cite{Nu01}),
$\Delta E_{\nu}=1, \Delta E_{\pi}=0$ (neutron shell gap reduced by 1 MeV from 
the Ref. \cite{Nu01} value),
$\Delta E_{\nu}=1, \Delta E_{\pi}=1$ (both neutron and proton shell gaps reduced
by 1 MeV from the Ref. \cite{Nu01} values), and
$\Delta E_{\nu}=1, \Delta E_{\pi}=3$ (neutron gap reduced by 1 MeV and proton gap
reduced by 3 MeV from Ref. \cite{Nu01} values).  The top panel is the inclusive
cross section for the two-proton knockout reaction from $^{44}$S (with the 
present experimental result shown as the dashed line and the experimental 
uncertainty cross-hatched).  The middle panel is the energy of the 
$2_1^+$ state, and the bottom panel is the reduced electromagnetic matrix 
element $B(E2;2_1^+ \rightarrow 0_{gs}^+)$.}
\end{figure}

\begin{figure}
\epsfig{file=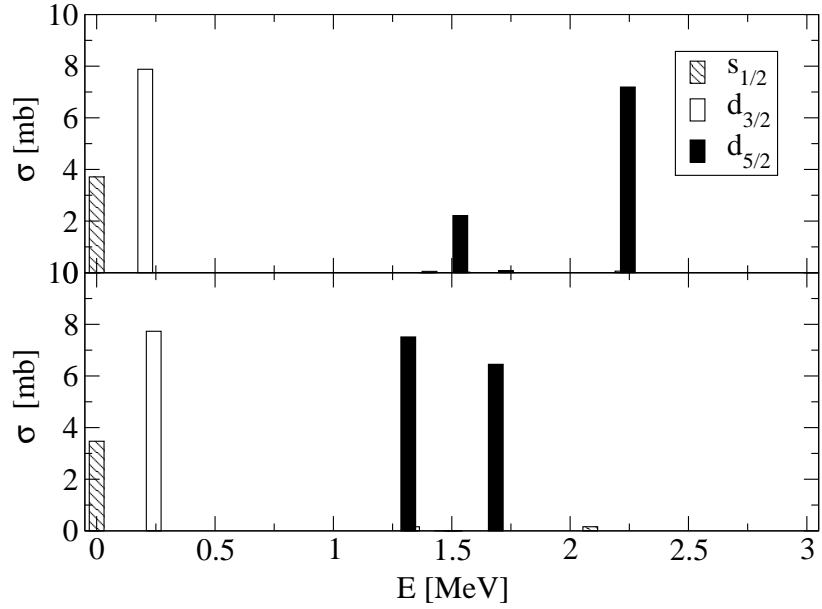,height=5in,angle=270}
\caption{Distribution of the strength for the $s_{1/2}$, $d_{3/2}$ and 
$d_{5/2}$ proton strength in $^{43}$P from the one-proton knockout
reaction on $^{44}$S calculated using two different values for the
$d_{3/2}-d_{5/2}$ proton spacing and (otherwise) the parameters from
Ref. \cite{Nu01}.  The top panel uses the spacing from Ref. \cite{Nu01}, while
the bottom panel uses a spacing 1 MeV smaller.}
\end{figure}

\end{document}